\documentclass[aps,pre,preprint,showpacs,floatfix]{revtex4}

\usepackage{graphicx}

\usepackage{amsmath}

\begin{document}

\title{Coarse Graining RNA Nanostructures for Molecular Dynamics Simulations}

\author{Maxim Paliy}
\affiliation{$M^2$NeT Lab, Wilfrid Laurier University  
75 University Avenue West
Waterloo, ON, Canada, N2L 3C5, http://www.m2netlab.wlu.ca/
}

\author{Roderick Melnik}
\affiliation{$M^2$NeT Lab, Wilfrid Laurier University  
75 University Avenue West
Waterloo, ON, Canada, N2L 3C5, http://www.m2netlab.wlu.ca/
}

\author{Bruce A. Shapiro}
\affiliation{Center for Cancer Research Nanobiology Program
National Cancer Institute, Frederick, MD 21702}

\date{\today}

\begin{abstract}
A series of coarse-grained models have been developed for the study of the 
molecular dynamics of RNA nanostructures. The models in the series  have one to three beads per nucleotide 
and include different amounts of detailed structural information. Such a treatment allows us to reach, 
for the systems of thousands of nucleotides, a time scale of microseconds 
(i.e. by three orders of magnitude longer than in the full atomistic modelling) 
and thus to enable simulations of large  RNA 
polymers in the context of bionanotechnology. We find  that the 3-beads-per-nucleotide models,
described by a set of just a few universal parameters, are able to describe different RNA conformations and 
are comparable in structural precision to the models where detailed values of the
backbone P-C4' dihedrals taken from a reference structure are included.
These findings are discussed in the context of the RNA conformation classes. 


\end{abstract}


\maketitle

\section{\label{intro}Introduction}

\begin{figure}
\includegraphics[width=3.25in]{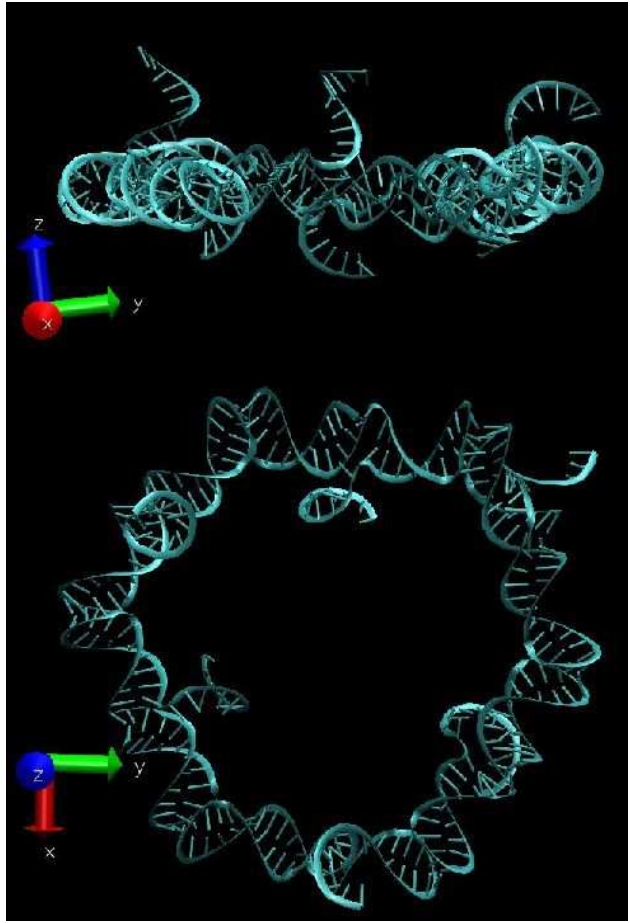}
\caption{\label{fig_ini} Side and top  views of the RNA nanoring structure in the ``new cartoon''  representation of VMD.}
\end{figure}

\begin{figure}
\includegraphics[width=3.25in]{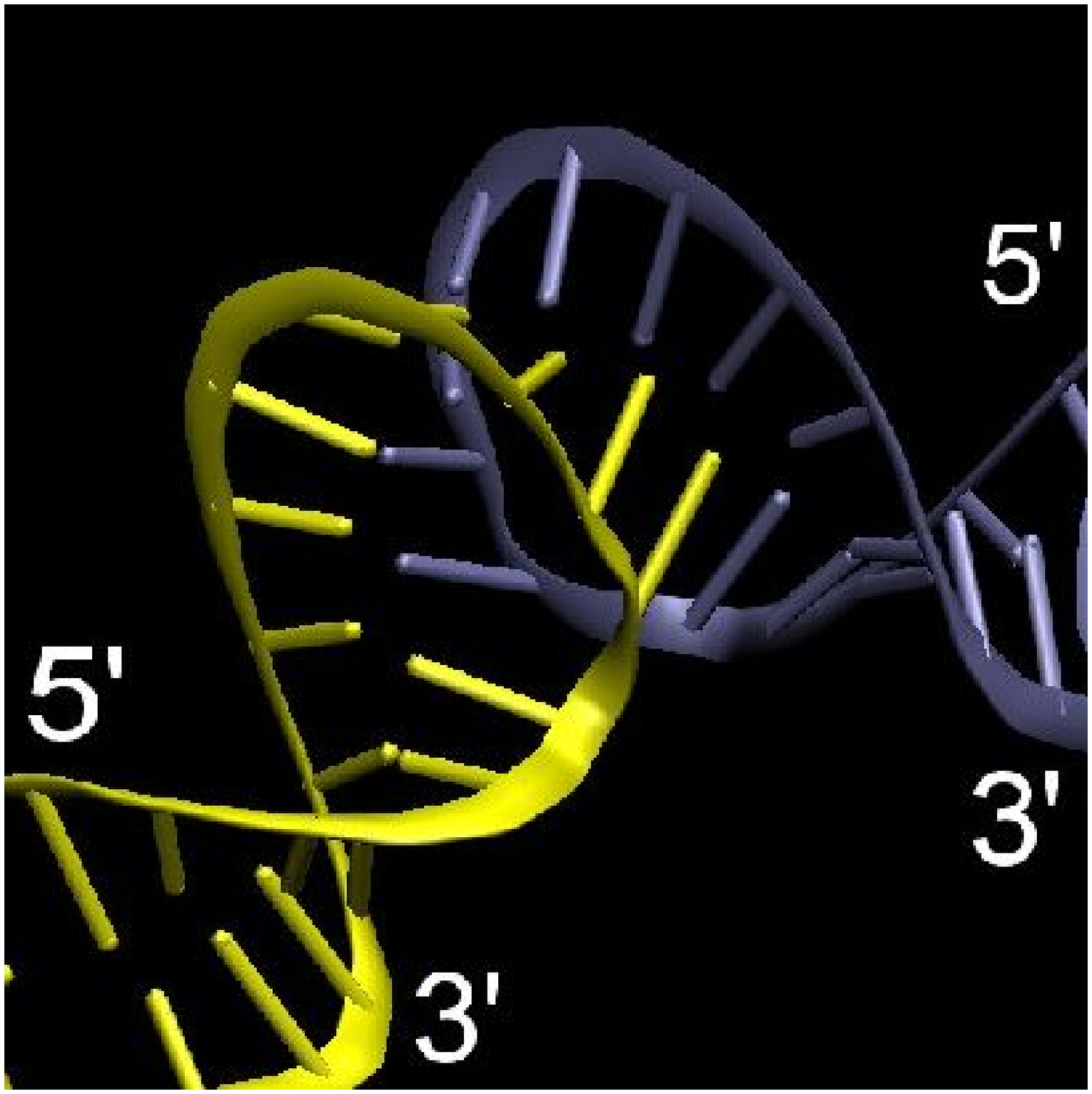}
\includegraphics[width=3.25in]{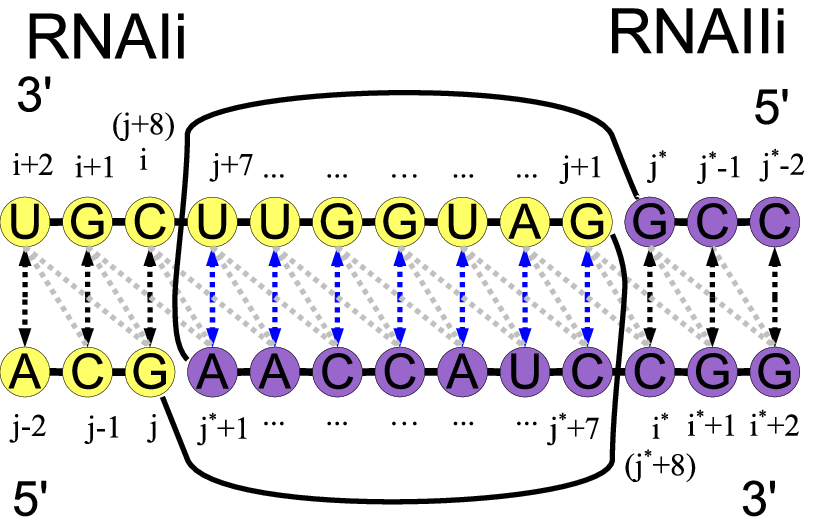}
\caption{\label{fig_snaphair}
Top: The ``new cartoon''  3D representation of one kissing loop corner of the RNA nanoring illustrating the base pairing 
and stacking. Bottom: 2D secondary structure sketch of the same kissing loop (colors correspond to the 3D view above); 
the main base pair bonds are denoted with the  arrows (black in the double-helical stems, blue in the kissing loop); 
two auxiliary bonds (Eq.~\ref{BPenergy}) per base pair are shown with gray dashed lines; 
the indexing scheme used in the simulation is shown; the base pairs are ordered in the sense of  stacking, 
that occurs continuously troughout the kissing loop; two long stretched lines connecting consecutive nucleotides 
correspond to two sharp kinks of the nucleic backbones visible in the 3D view above.}
\end{figure}

Recent progress in  understanding RNA structure brought to light a new concept of RNA architectonics - a set of
recipes for (self-)assembly of  RNA nanostructures of arbitrary
size and shape \cite{JaegerChworos,Jaeger}.  Smallest RNA  building blocks  - ``tectoRNAs'',
typically bearing well-defined structural features,  such as
``right angle'' \cite{JaegerChworos}, ``kink-turn''  \cite{Jaeger,Holbrook}  or  
``RNAIi/RNAIIi'' \cite{YingLing} motifs were manipulated, 
either experimentally \cite{JaegerChworos,Jaeger} or
via a computer modelling \cite{YingLing}, into the desired  2D or 3D nanostructures
(squares, hexagons, cubes, tetrahedrons {\em etc.}) that can be further assembled
into periodic or quasi-periodic patterns. Compared to DNA nanostructures,
RNA as a nano-engineering material brings additional challenging features, such as much larger diversity in 
tertiary structural building blocks [$\sim 200$ versus $\sim 20$  for DNA \cite{Jaeger}] and, often, 
increased conformational flexibility [see e.g. \cite{RNADNAbook},  p. 320].
Most useful insights about the behavior of the above-mentioned nanostructures  
can be gained by  all-atom Molecular Dynamics (MD) simulations \cite{RNADNAbook}. 
However, presently, the time scales that can be achieved in the all-atom MD amount to a few (tens) nanoseconds only, 
which is by many orders of magnitude less than the duration of the slowest processes 
occurring in biomolecules (micro- to milli- to seconds).
For example, in a recent study \cite{PhysBio2009}, we analyzed, via all-atom MD
simulation, a simple  RNA nanostructure of about 13 nm in size (330 nucleotides), a hexagon-shaped  
RNA ring \cite{YingLing}, termed ``nanoring'' in what follows. It is composed (\ref{fig_ini}) of six RNAIi/RNAIIi 
complexes, which make up its sides, type A double helices, joined by the ``kissing loop'' motifs at the corners 
(e.g. AACCAUC septaloop  is paired with UUGGUAG loop). ~\ref{fig_snaphair} shows  
the patterns of base pairing and stacking in the kissing loops.

In order to reach at least microsecond time scales in simulations of such nanostructures 
(hundreds to thousands nucleotides in size), one needs to consider a coarse-grained (CG) treatment, where the groups of atoms 
are represented by the CG interactions centres - ``beads'', and effective interactions between such beads are set in a way 
to fit the nanostructures atomic connectivity, thermal, mechanical properties {\em etc}. 
Two kinds of data are often used in the fitting process:
(i) the experimentally available  structural information as well as other known properties of interest 
(which can be limited and/or incomplete), and (ii) a host of very detailed atomistic data obtained from 
all-atom MD simulations. 
Namely, the  parameters for  a CG model can  be derived from both experimental and full-atom MD data by Boltzmann Inversion (BI)
 \cite{Bzinversion} of the Radial Distribution Functions (RDFs), using the Inverse Monte Carlo scheme 
\cite{Lyubartsev1995} or, in the case of all-atom MD simulation only, 
with the ``Force Matching'' method \cite{EA1994} [for some recent approaches to biomolecules, see e.g.
 \cite{gregCGbook,Voth,Noid2008,Noid2008a,Tozzini,Tozzininew}]. 
Finally, such a  CG model can be further investigated using Coarse-Grained Molecular Dynamics (CGMD), 
which allows one to reach much longer time scales [although the dynamics 
of the original system is not always adequately represented \cite{Voth}]. The main challenge is  to describe the RNA on a coarse-grained level 
with just a few universal parameters, thus adopting the strategy of a ``CG forcefield'' (FF).
This proved to be a difficult task, in particular in the sense of transferability of such a CG model to other structures,
not used in the fitting. Transferability problems are notorius even for condensed matter, and it is even more true for the biomolecules, 
that show enormous structural diversity, see e.g. \cite{gregCGbook,Ouldridge2009,Ghosh2007,Nielsen2004}. 
In the latter case, instead of a FF approach, a lot of detailed structural information 
(such as equilibrium values of bonds, angles, dihedrals, nonbonded interatomic distances from the experimentally resolved structures) is supplied to the CG model, thus providing its  structural precision, although in describing a given structure only.
Such a structurally biased approach is often termed the  Self Organized Polymer, SOP \cite{Thirumalai3}. 

In the CG modelling of RNA, several recent advances should be mentioned 
(see e.g. \cite{Thirumalai,Thirumalai2,Thirumalai3,Trylska,Jonikas,Pincus2008,CAO2005,CAO2009,Tan2006,Das2007,Parisien2008,DokholyanRNA,Gherghe2009}). 
Besides, many of the RNA modelling approaches focusing on the 3D tertiary structure prediction [reviewed in \cite{Shapiro2007}] 
make use of some coarse-graining via variety of methods, such as building complex CG energy functions and use of 
pre-compiled databases of known motifs. However, comparatively few of them are able to produce long-time dynamics once the 
3D structure is known. Existing CG models for RNA, similar to those for DNA \cite{Knotts2007,Sambriski2009a,Ouldridge2009}, 
include one to three coarse grained units (beads) per nucleotide. One-bead RNA models normally use the SOP approach to 
achieve the  desired structural precision \cite{Thirumalai,Thirumalai2,Trylska,Jonikas} [recent model developed in 
\cite{Jonikas} uses special bonds to fix the tertiary contacts only]. On the other hand,
the most recent three-bead RNA CG model \cite{DokholyanRNA,Gherghe2009} achieves a reasonable success
in structural precision (for up 100-nucleotide chains) even with a CG FF approach, 
that includes complex sequence-dependent interaction details and many-body treatment of the stacking, 
base-pairing and hydrophobic interactions [for some larger structures, it still needs 
the special artificially introduced bonds to fix long-range tertiary contacts \cite{Gherghe2009}]. 
However, the technique employed to simulate the evolution of the system, Discrete Molecular Dynamics \cite{DokholyanRNA}, 
while providing a means for fast folding, at the same time necessarily dictates the use of simplified step-wise 
potentials, which may introduce additional sources of imprecision into the model. By contrast, we use
continuous potentials to describe the energetics of RNA, which should make the dynamics of the system more realistic.
To put our RNA CG modelling efforts better in the perspective of the previous studies, we focus on  the development of a simplest
continuous CG energy function, that enables us to do long-time CGMD simulations of the huge aggregates such as RNA nanostructures 
(this simplicity comes at an expense - e.g. presently the secondary stucture should be known to our models, a requirement 
that will be lifted in the future).

Namely, in the present paper we explore the avenues from the SOP approach to a RNA CG forcefield approach by 
varying number of beads and amount of atomistic structural information in our series of models. 
We show that the inclusion of just the dihedral pseudo angles P-C4' in a SOP manner brings about the same structural precision 
to the model, as the full SOP approach. Besides, a simple modification of the dihedral angle terms to allow an alternative 
value of each dihedral can render even the RNA FF model sufficiently precise to describe the studied 
RNA nanoring. These findings are consistent with the existence of the RNA conformation classes, 
based on P-C4' dihedrals \cite{RNAtorsional}. 

\section{\label{model} Coarse-Grained Model}

The development of a CG model consists of two major stages: 
(i) choice  of the groups of atoms to be combined in a single CG bead, 
and (ii) selection  of the functional forms and fitting of the parameters for the 
effective interactions between the beads. 

In the case of  nucleic acids, the simplest choice for stage (i) is one bead per nucleotide 
(beads being placed normally on the phosphate groups) which allows one to use experimentally available structural 
data \cite{Trylska,Tozzininew,Jonikas}. 
However, as it has been recently shown in \cite{RNAtorsional}, the available experimentally RNA conformations 
may be described well by just two torsional angles, between the P  and C4' atoms. Connecting the beads placed 
at the C4' atomic sites with the base pairing bonds is not suitable in terms of the geometry, 
as such bonds are too far off the axis of the double helix. Instead, we adopt a representation with three beads per nucleotide, 
that corresponds to the (P)hosphate, (S)ugar, and nucleic (B)ase, 
respectively, which is a natural choice  for nucleic acids. 
It has already been used in a few recent articles both on DNA \cite{Knotts2007,Sambriski2009a} and on RNA
\cite{DokholyanRNA,Gherghe2009,CAO2009}.  Note that to exploit the idea of RNA conformation classes 
we chose to place beads on the existing atoms rather then on the center of masses of groups of atoms. 

\begin{figure}
\includegraphics[width=1.61in]{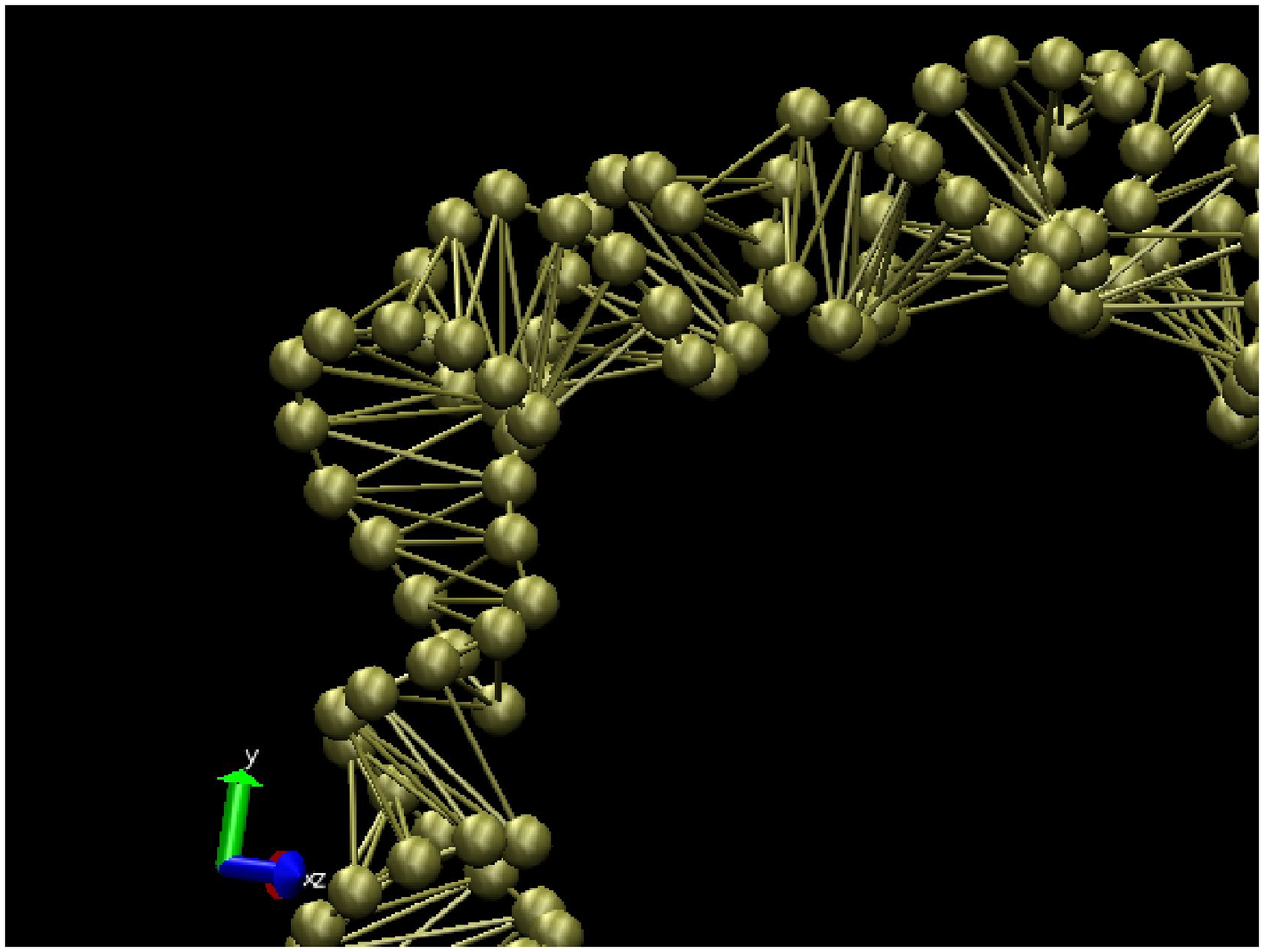}
\includegraphics[width=1.61in]{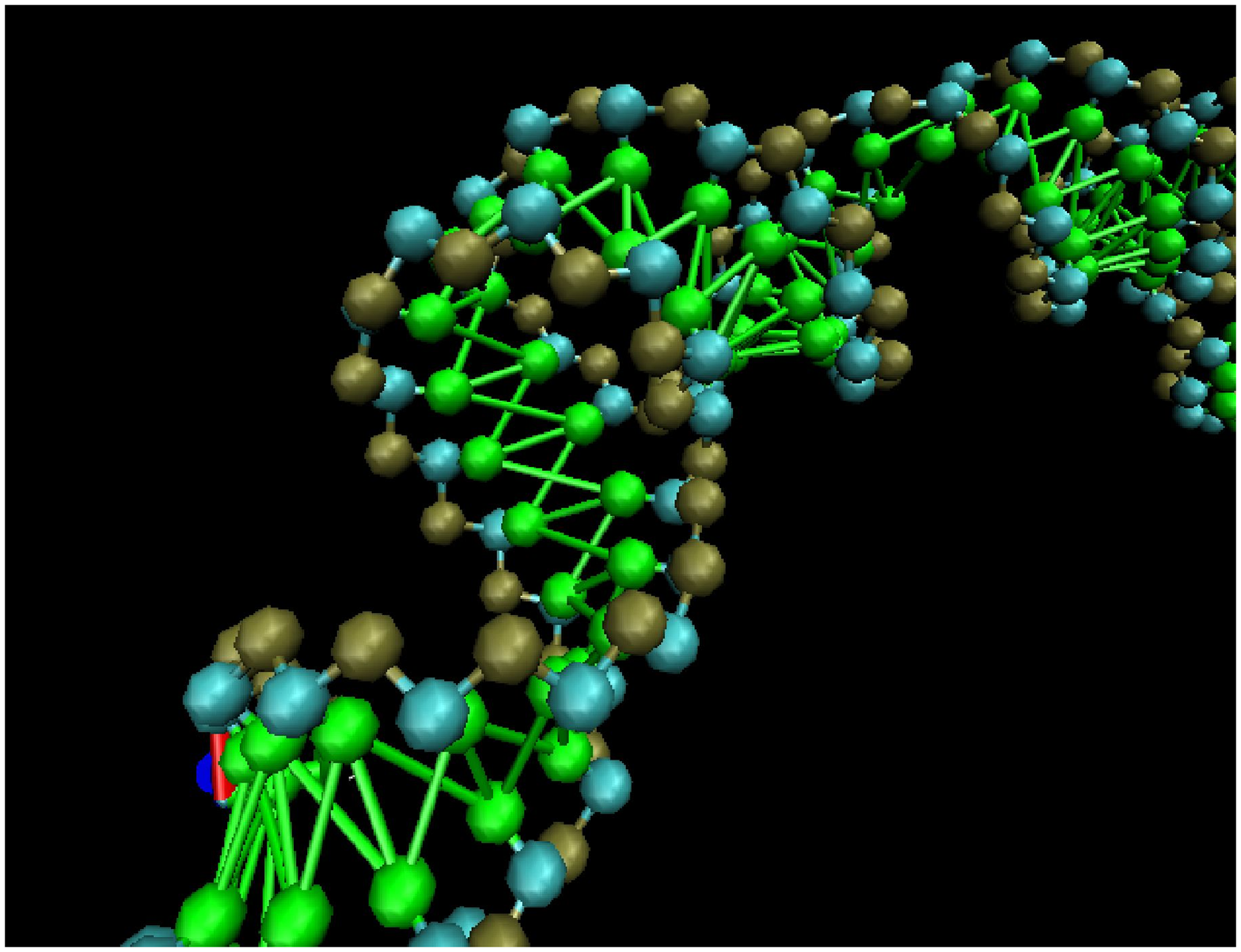}
\includegraphics[width=3.25in]{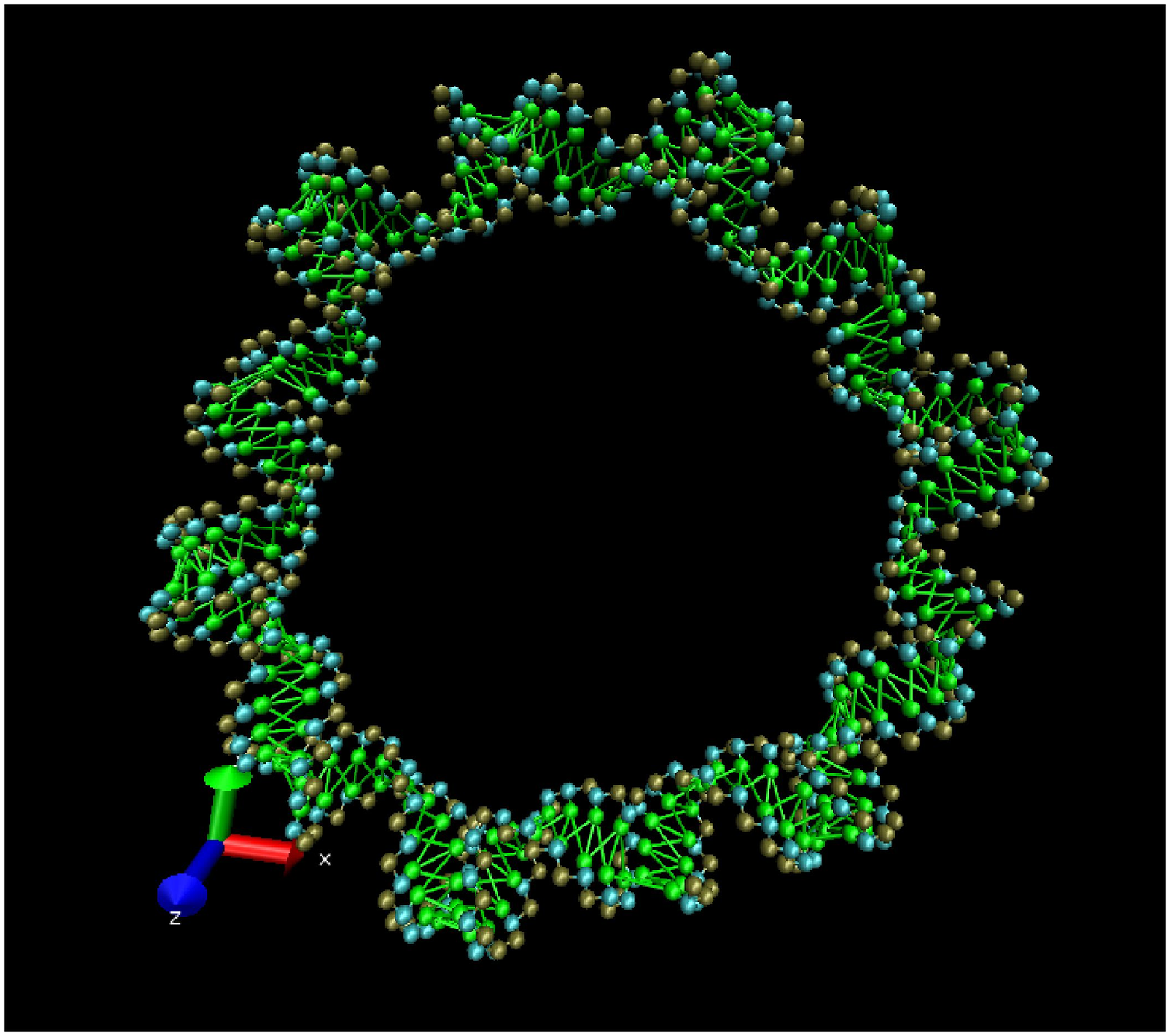}
\caption{\label{fig_snap123} 
CG representations for the RNA nanoring shown in Fig.\ref{fig_ini}.
From top to bottom: Zoomed views of one "kissing loop" in 1B and 3B representations, and the 
full RNA nanoring 
in the 3B representation. The phosphate (P) beads are shown in brown, 
the sugar (S) beads - in cyan, the base beads (B) -  in  green. 
The bonding  of the backbone and between the bases is shown with the lines.}
\end{figure}

The sample configurations of the RNA nanoring in on the one- and three-bead variants of our models 
(denoted as 1B and 3B in what follows) are depicted in the \ref{fig_snap123}. In the 1B case, all beads of the single 
type (with the mass  $m^{(P)}=321.5$~a.m.u.)  are placed on the P atoms in the phosphates. In the 3B case, two types of 
beads are thus placed on the P atoms and C4' carbons, while a number of plausible choices are possible for the 
placement of the third base bead. We found the following variant to be most convenient: 
N9 atom of purines and N1 atom of pyrimidines. The masses of the beads in the 3B representation are taken as 
$m^{(P)}=109$~a.m.u. , $m^{(S)}=120$~a.m.u.,  $m^{(B)}=92.5$~a.m.u. 

In our model, the beads are organised into several single chains, that correspond to the basic building blocks 
of the studied nanostructure, and the connectivity inside which is never broken in the course of  simulation. 
For example,  the nanoring from \ref{fig_ini} is built up from six chains that form the sides of a hexagon,
and are folded into double helical stems with septaloops at both ends. Dangling 5' and 3' ends of 
the chains, found in the middle of the hexagon sides, are excluded from the CG model in order to focus on the 
core of the nanoring (264 nucleotides). The total interaction energy has the following form:
\begin{equation}
\label{CGenergy}
V = V_{conn} + V_{bp} + V_{nb},
\end{equation}
with the standard  chain connectivity contribution  $V_{conn}$:
\begin{equation}
\label{CONNenergy}
V_{conn} = \sum_{chains} \left( \sum _{bonds} V_{b}(r-r^{(0)}) + \sum _{angles} V_{a}(\theta-\theta^{(0)}) 
+ \sum _{dihedrals} V_{d}(\phi-\phi^{(0)})  \right) , 
\end{equation} 
where  $V_{b}(r)$, $V_{a}(\theta)$, $V_{d}(\phi)$
are intra-chain terms that correspond to the  energies of bonds, angles, and dihedrals 
(often abbreviated ``b/a/d'' in what follows), while $r^{(0)}$, $\theta^{(0)}$, $\phi^{(0)}$ 
are the  equilibrium  values for b/a/d, respectively. 

The energy term $V_{bp}$ accounts for the interactions between the base pairs.
In the case of the nanoring, these include the contributions from the base pairs found inside  the double-helical part 
of a single chain, as well as between those septuplets of the base pairs belonging to different chains, 
that form the kissing loops. Following the idea of \cite{Tozzininew}, we express the base pair interactions 
with  three bonds per base pair instead of one:
\begin{equation}
\label{BPenergy}
V_{bp} =  \sum _{i,j \in (base pairs)} U_{i,j}(r_{i,j} - r^{(0)}_{i,j}) 
+  U_{i+1,j}(r_{i+1,j} - r^{(0)}_{i+1,j}) +  U_{i+2,j}(r_{i+2,j} - r^{(0)}_{i+2,j}),
\end{equation}
which lets us  enhance structural accuracy, since it takes into account both the hydrogen bonding between bases and the stacking interactions.
As illustrated in \ref{fig_snaphair}~(bottom, arrows and gray dashed lines), 
a base $j$ interacts not only  with its counterpart $i$, but also with the neighboring bases $i+1$, and $i+2$ 
from the anti-sense part of the same double-helical stem (if present). A special care is required for laying out the base 
pair interactions in the kissing loops. Since a base paired  kissing loop section closely resembles the double-helical structure, 
and  the stacking occurs continuously from one stem helix through the loop–-loop helix to the other stem helix
\cite{LeeCrothers}, in the RNA nanoring all the bases have three interacting neighbours 
(possibly from two different chains). The connectivity between the base pairs is thus maintained throughout the 
time evolution.

The remaining energy contribution $V_{nb}$ corresponds to the interactions between all bead pairs 
not involved in the bonded interactions described above. It has the following form:
\begin{equation}
\label{NBenergy}
V_{nb} =  \sum _{i,j \in (nonbonded)} v(r_{ij}).
\end{equation} 
In the  present paper we take the simplest Weeks-Chandler-Andersen (WCA) form \cite{Weeks1971} 
for the nonbonded potential $v(r)$. It consists of the repulsive part of the Lennard-Jones potential, 
and expresses the steric repulsion between the beads via energy $\varepsilon$ and  bead diameter $\sigma$. 

The MD for the CG model is implemented via the DL\_POLY~$2.19$ package \cite{dlpoly}. 
The resulting CG model is relaxed via an energy minimization (conjugate gradient method) and then equilibrated 
at a constant temperature in the NVT ensemble [Evans algorithm, \cite{EvansThermo}] with open boundary conditions and 
the time step  of $0.01$~ps, sufficiently small to conserve the energy 
of the system in the constant energy runs. The cutoff of $10$~\AA \  is applied to  nonbonded 
interactions. Visualization and data processing of both all-atom MD and CGMD simulations are carried with VMD \cite{vmd}, 
using in-house developed scripts.

\section{\label{fitting} Fitting of the CG Parameters}

The total energy of the CG model, Eq.~\ref{CGenergy}  thus contains a number of parameters 
for the b/a/d, base pair terms (as well as those for the nonbonded interactions).
Our general approach to the fitting of these parameters is the following:
(i) the histograms of the values of various bonds, angles and dihedrals, as well as 
the Radial Distribution Functions (RDF) between all sorts of beads
are extracted from all-atom MD trajectories; (ii) these distributions are used 
to fit the CG model parameters via the Boltzmann Inversion method \cite{Bzinversion}.
Namely, given a probability distribution function $P(q)$ for a degree of freedom $q$, the corresponding potential 
of mean force (PMF) $V_{eff}(q)$ is determined via the following formula:
\begin{equation}
V_{eff}(q) = -k_B T \ln(P(q)).
\label{BZinv}
\end{equation}
Note that thus obtained $V_{eff}(q)$ coincides with the true potential energy  only for the case of  
a single degree of freedom $q$, and generally it can serve only as a first approximation used in  
a subsequent iterative procedure, which may not always be successful, because many variables have to be fitted 
simultaneously. Fortunately, different energy contributions usually show a certain hierarchy, which allows their 
refinement in succession, in order of their decreasing strength, e.g. $V_{bond} \rightarrow V_{angle} \rightarrow V_{van-der-Waals} \rightarrow V_{dihedral}$ \cite{Bzinversion}. 

We fit the effective potentials for the bonded degrees of freedom $V_{eff}(q)$, Eq.~(\ref{BZinv}), 
by their Taylor expansions (up to quartic) around their global minima:
 \begin{equation}
V_{eff}(q) = \frac{k}{2} (q-q^{(0)})^2 + \frac{k'}{3} (q-q^{(0)})^3 +\frac{k''}{4} (q-q^{(0)})^4. 
\label{Veffquartic}
\end{equation}
We thus obtain  an equilibrium value $q^{(0)}$ of a bonded degree of freedom $q$ and the coefficients 
$k$, $k'$, $k''$. For the 1B CG model, we used all three terms, 
while for the 3B CG model the use of only the harmonic term $k$ proved to be sufficient for 
our purposes (for the latter case, we made some tests  with the anharmonic coefficients $k'$ $k''$ included, 
and found no important differences; they may be useful in the future, for the fitting of the CG model
to the dynamical properties, such as diffusivity). We find that for most of the bonded degrees of freedom (in the 3B case) 
the initial values of the parameters obtained  directly from the Eqs.~(\ref{BZinv}) and (\ref{Veffquartic}), already reproduce 
sufficiently well the  histograms in the CGMD simulations, so that only seldom subsequent adjustments 
are required. They are done by  manually introducing small changes to the coefficients in  Eq.~\ref{Veffquartic}, 
in order to improve matching between the MD and CGMD distributions. 
For further refinement of the model, we plan to resort also to a more systematic 
fitting procedure, involving the iterative procedures and Force Matching method \cite{EA1994} with cubic spline potentials
\cite{Voth} in particular.

We used two all-atom MD data sources: (i) a $6$~ns  $300$~K trajectory of a simple RNA double A-helix dodecamer (GCGCUUAAGCGC);
(ii) a $2$~ns  $310$~K trajectory of a complex RNA nanostructure - nanoring. 
Both systems have been simulated in the explicit water with Mg and Na counter ions. 
Further details about these MD runs can be found in the Supporting Information. 
While the data derived from the dodecamer served us as a source for ``double-helical'' parameters 
(due to a longer trajectory they are also more reliable), the set of data derived from 
the nanoring, allowed us to introduce ``non-helicity'' into the model in a controllable manner.

It is important to stress that, for the RNA nanoring, the above-mentioned degrees of freedom are not distributed according 
to Boltzmann statistics only, but their distributions also reflect the spatial inhomogeneity of 
the system. Therefore, an attempt to represent such a degree of freedom via a single potential function using the BI method 
would lead to instability of the desired structure in the CG model, because such a degree of freedom  would be 
discriminated against energetically in certain regions. Instead, one may introduce some local modifications to the potential 
functions. The ultimate strategy of this sort is the SOP approach, where each instance of such a degree of freedom has its 
own equilibrium value depending on its location in the molecule. 

Thus, we consider three different parameter sets: 
(i) ``SOP'' parameter set, where the coefficients $k$, $k'$, $k''$ of the potential functions $V_{eff}(q)$
 are uniform throughout the system, while the equilibrium values $q^{(0)}$
are unique for each instance of the b/a/d;
(ii) ``SOP-dihedrals'', where the SOP approach (i) is applied  only to the dihedrals of the nucleic acid backbone, 
but not to the bonds, base pairing bonds or angles, and 
(iii)  ``forcefield'' (FF) parameter set where each instance of a degree of freedom is  described 
by all uniform parameters (including their equilibrium values) throughout the system. 
In all cases, for the uniform part, we used the CG parameters $k$, $k'$, $k''$, and $q^{(0)}$ 
extracted from the dodecamer data. However, for the SOP and the SOP-dihedrals parameters sets, we replaced the uniform 
equilibrium  values with the full (inhomogeneous) sets found either in the initial non-equilibrated structure of the nanoring 
(we term such configuration ``ini1'' in what follows), or by the averaging of each instance of an equilibrium 
value over the all-atom MD trajectory of the nanoring (``ini2'').

\section{\label{results} Results}

\subsection{\label{1Bfitting} 1B representation of the Coarse-Grained Model}

The simplest  1B representation of the model includes the following CG degrees of freedom: 
$(P_{i}P_{i+1})$ backbone bonds,  $(P_{i}P_{i+1}P_{i+2})$ backbone angles and $(P_{i}P_{i+1}P_{i+2}P_{i+3})$ backbone dihedrals, 
three base-pairing bonds $(P_{i}P_{j})$ , $(P_{i+1}P_{j})$,  $(P_{i+2}P_{j})$ per ($i,j$) base pair, 
as well as the distances for the nonbonded $(PP)$ pairs. The indices $i$ and $j$ denote the nucleotide numbers in the sequence  
(counting from the 5' end), and they are omitted in what follows wherever possible without the loss of clarity.

\begin{figure}
\includegraphics[width=3.25in]{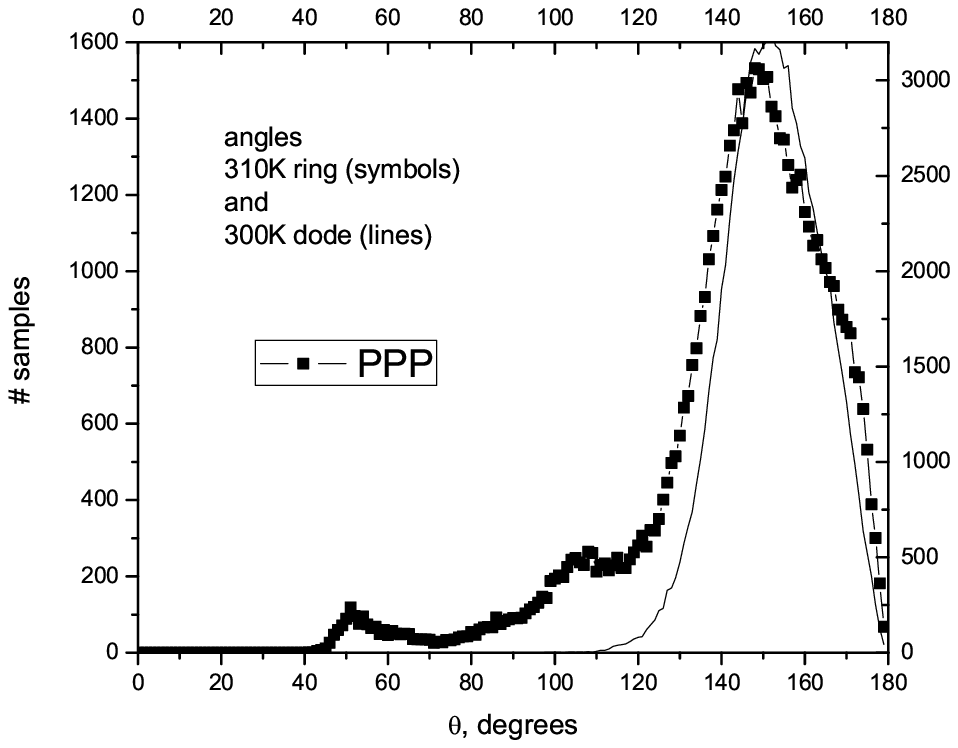}
\includegraphics[width=3.25in]{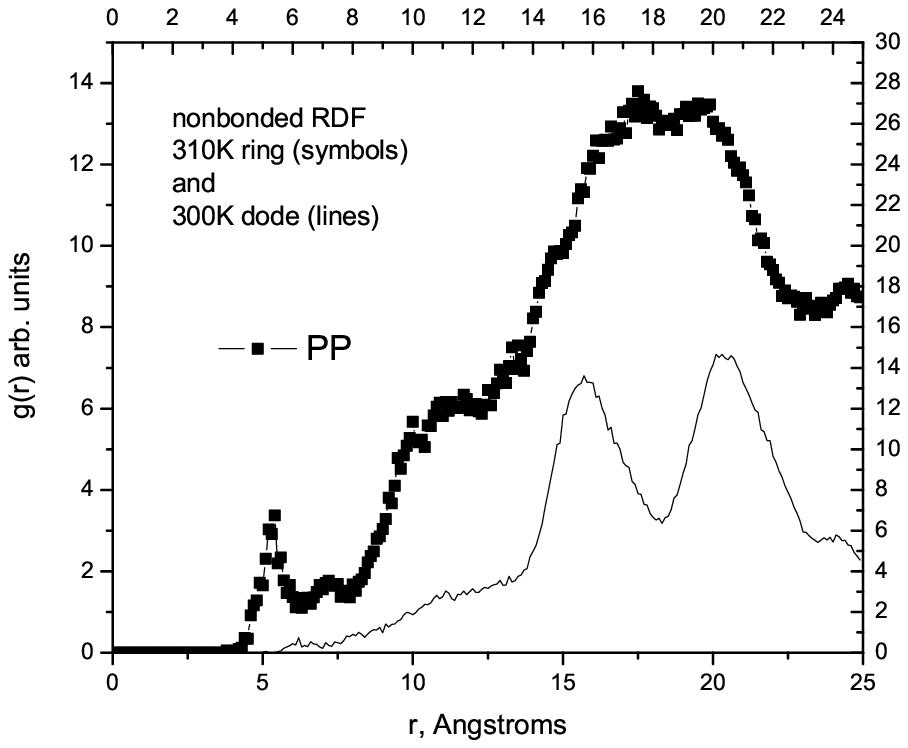}
\caption{\label{fig_1b_mdall}
Angle histograms and nonbonded RDFs from all-atom MD runs for the 1B representation of the model.
The data for  the RNA nanoring are shown with symbols, while those for  the RNA dodecamer are 
shown with thin lines.}
\end{figure}

The histograms of the backbone angles as well as RDFs for the nonbonded pairs 
from  all-atom MD  data are plotted in \ref{fig_1b_mdall} for both the RNA dodecamer and the RNA nanoring in comparison 
(data for all the 1B CG degrees of freedom can be found in the Supporting Information, Fig.~S1). 
The distributions for  the RNA nanoring are in general broader, they contain extended tails, that reflect the existence 
of nonhelical regions. While the distribution of the backbone and base pair bond lengths are more simple and unimodal, 
for both the dodecamer and the nanoring, the remaining distributions (angles, dihedrals, and nonbonded pairs) 
for the nanoring show a number of fine features, absent in the dodecamer, and taking into account of which is crucial 
for the development of a CG model. Namely, the angular distributions for the nanoring show additional small peaks at $\approx 60^{\circ}$ and   $\approx 110^{\circ}$ besides the main peak at $155^{\circ}$, the dihedrals show additional broad peak  at  $\approx  - 160^{\circ}$ besides the main peak at $14.4^{\circ}$, and the nonbonded RDF shows a small peak at 
$\approx5$~\AA  \  due to the closely spaced phosphates. As close examination  of the atomic configurations 
reveals, these features are associated mostly with the regions of the kissing loops.

Since it does not make sense to fit multi-modal distributions of the bonded CG degrees of freedom with 
simple potentials of Eq.~(\ref{Veffquartic}), in the 1B variant of the model we considered only SOP parameters sets 
with the coefficients  $k$, $k'$, $k''$ derived from the dodecamer data.
Besides, as the peak of the nonbonded RDF at  $\approx5$~\AA  \ dictates  that the nonbonded interaction potential 
does not discriminate such small interbead spacings energetically, we have taken the values $\varepsilon = 0.1$~kcal/mol, $\sigma = 5.0$~\AA \  for the nonbonded WCA parameters. Table~S1 lists the working values of the parameters for the 1B CG model. 
We subjected the resulting 1B CG representation of the RNA nanoring to $500$~ns equilibration 
at $T=300$~K. Various distributions from these equilibration runs  are plotted in Fig.~S2 together 
with the analogous all-atom MD data for comparison. While a more detailed account can be found in the Supporting 
Information, here we emphasize the main result - the 1B CG model fails to reproduce the overall 
shape of the nanoring, which collapses to various unrealistic configurations, characterized by the
abundance of too closely spaced nonbonded beads, Fig.~S2.
This is the consequence of the above-mentioned $\approx5$~\AA  \ restriction on the repulsive nonbonded potential 
(while the all-atom MD nonbonded RDFs suggest that the beads should be at least $\approx 10 - 15$~\AA \ in diameter). 
It is possible to introduce a more complex  nonbonded pair potential, with multiple minima.
However, this opens a number of questions about the relative depths of 
the minima/heights of the barriers and the existence of unrealistic spurious configurations in 
the result. While a complex pair potential can be represented with splines and adjusted in a 
very detailed manner \cite{Voth}, this does not secure us from the latter caveat.  
Besides, for the FF parameter set in the 1B representation, one needs to introduce the angular and dihedral terms of the 
fairly complex shapes too. This combination of factors led us to abandon the 1B representation as unsuitable in favour of the 
3B representation.

\subsection{\label{3Bfitting} 3B representation of the Coarse-Grained Model}

The full list of  different energy terms for the 3B representation of the CG model includes the 
following (the layout of the bonding terms is shown
in \ref{fig_snaphair} and \ref{fig_snap123}). Along the  nucleic acid backbone the $(P_{i}S_{i})$, $(S_{i}P_{i+1})$, $(S_{i}B_{i})$ bonds 
 between the nearest neighbours, as well as  
$(P_{i}S_{i}P_{i+1})$, $(S_{i}P_{i+1}S_{i+1})$, $(P_{i}S_{i}B_{i})$, $(B_{i}S_{i}P_{i+1})$ angles,
$(P_{i}S_{i}P_{i+1}S_{i+1})$ and $(S_{i}P_{i+1}S_{i+1}P_{i+2})$ dihedrals are included. 
Along the base-paired parts of double helices and kissing loops 
the $(B_{i}B_{j})$ , $(B_{i+1}B_{j})$, and  $(B_{i+2}B_{j})$ bonds are included. 
Besides, due to the topology of the 3-bead nucleic backbone, we introduce dummy ``zero energy'' bonds 
between the nearest $(S_{i}B_{i+1})$,  $(B_{i}S_{i+1})$, and  $(B_{i}B_{i+1})$  
neighbours along the backbone in order to exclude thems from the nonbonded interactions, 
since these bonds are already restrained by the above-mentioned set of backbone terms.

\begin{figure}
\includegraphics[width=3.25in]{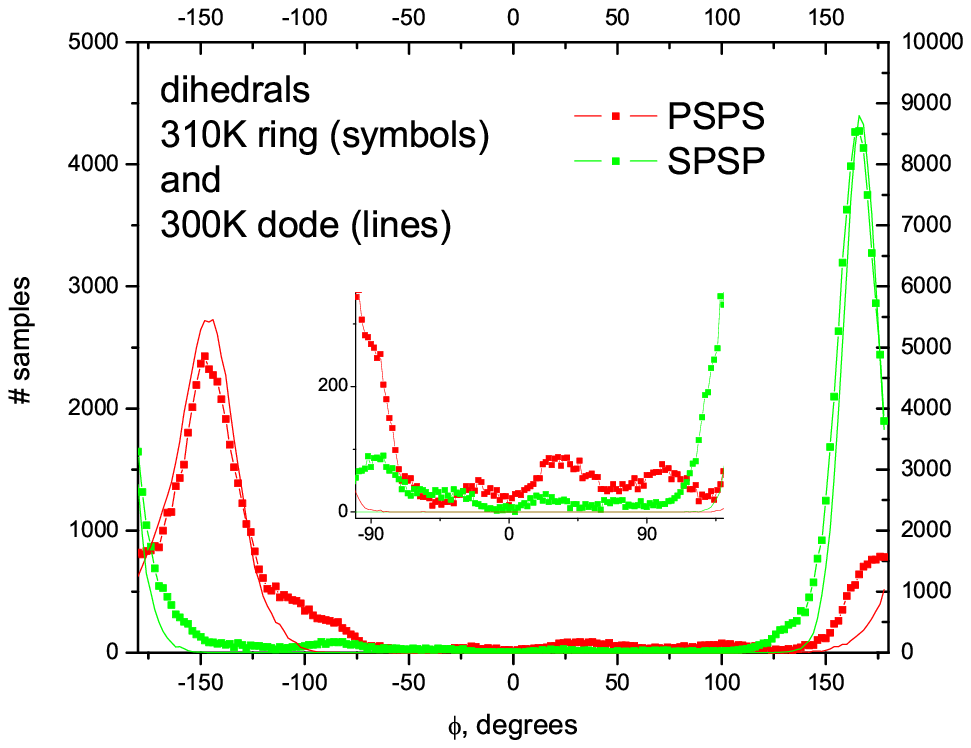}
\includegraphics[width=3.25in]{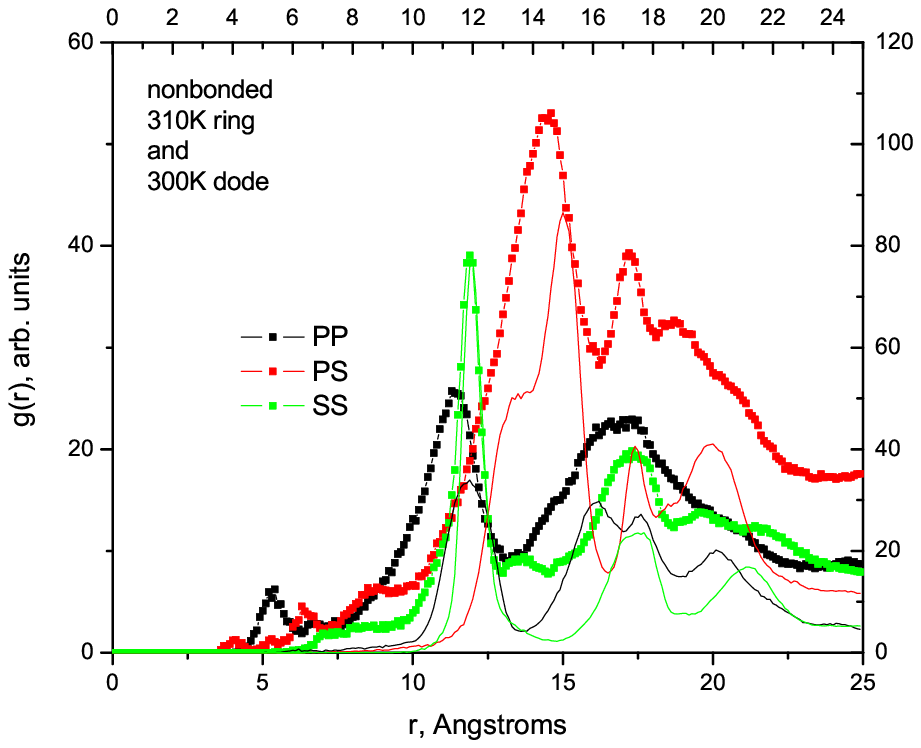}
\caption{\label{fig_3b_bad} 
Top: Histograms for  dihedral angles from all-atom MD runs  in the 3B representation of the model. 
The inset shows the zoomed portion of the figure near the baseline  in the interval $[-100^\circ , 140^\circ]$.
Bottom: Radial Distribution Functions for nonbonded interactions from all-atom MD runs 
for the 3B representation of the model. The data for the RNA nanoring are shown with symbols, 
while those for  the RNA dodecamer are shown with thin lines (the color coding is the same in both cases).}
\end{figure}

Our 3B  model includes also 6 different non-bonded  bead pairings $(PP)$, $(PS)$, $(PB)$, $(SS)$, $(SB)$, and $(BB)$. 
The RDFs from the all-atom MD runs for three selected pair types are shown 
in \ref{fig_3b_bad} for both studied systems (the full set of data for all  nonbonded as well as bonded
degrees of freedom can be found in Figs.~S3-S4). The nonbonded RDFs for the RNA nanoring 
(\ref{fig_3b_bad}) show well pronounced  peaks/tails in the interval between $5$~\AA \ and $10$~\AA , which are absent 
in the case of the dodecamer. As in the 1B case, these features, caused by closely spaced beads in the kissing loops,
dictate the choice of the same nonbonded parameters, $\varepsilon = 0.1$~kcal/mol, $\sigma = 5.0$~\AA.

The 3B bonded distributions for the RNA nanoring show much 
less pronounced fine features, compared to the analogous plots for the 1B case. 
In fact, all the bonded terms (except the dihedrals) have unimodal distributions closely 
resembling Gaussians, which allows us to retain the harmonic coefficients $k$ only in Eq.~\ref{Veffquartic}. 
More complex distributions are demonstrated by the dihedral angles. They are plotted in \ref{fig_3b_bad} for both 
the RNA nanoring and the RNA dodecamer for comparison. For example, the (PSPS) dihedrals contain  two shoulders near 
the main peak at $\approx -153.4^\circ$ consistent with two RNA conformational classes  \cite{RNAtorsional}, 
as further explained in the Discussion. 

\begin{figure}
\includegraphics[width=3.25in]{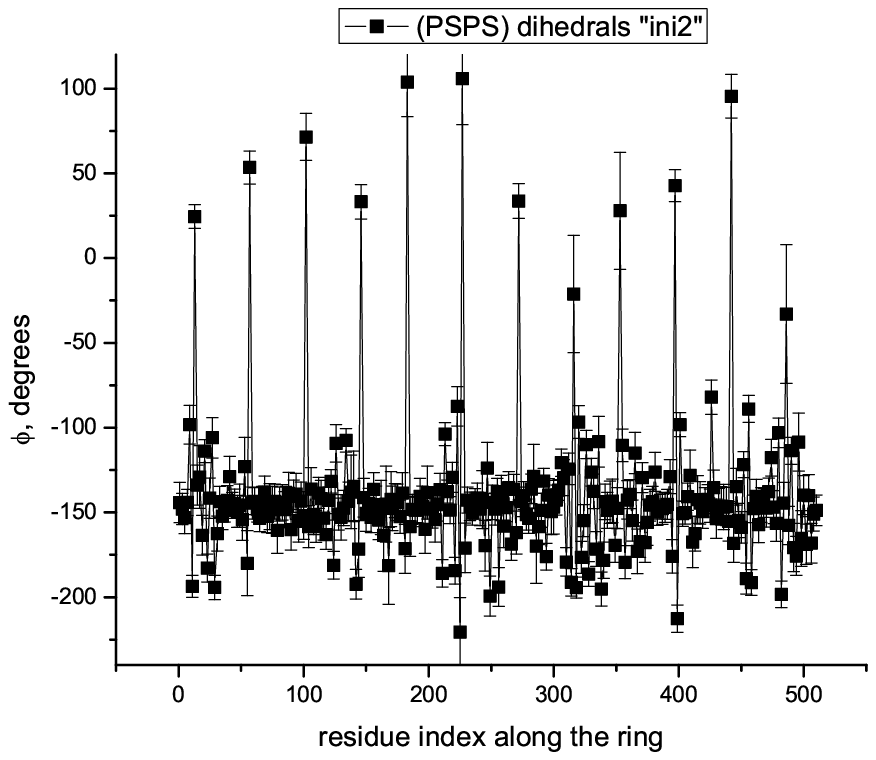}
\includegraphics[width=3.25in]{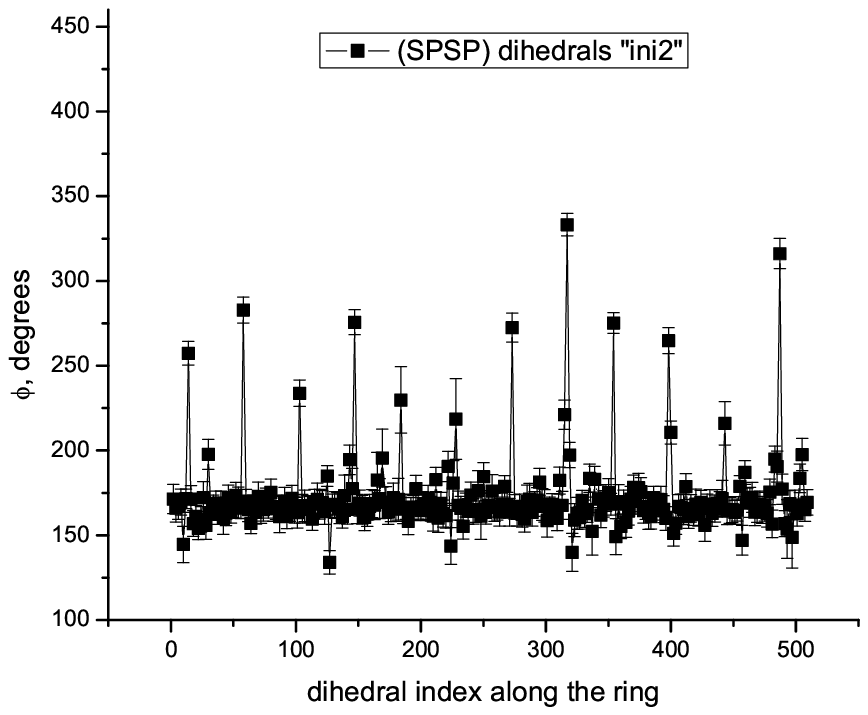}
\caption{\label{fig_dihfield} 
The  "SOP" fields  of the dihedral angles  introduced in the 3B model. Top: for initial  configuration termed "ini1". 
Bottom: averaged over a all-atom MD trajectory (termed "ini2").  To represent better the deviating dihedrals, the  plotting intervals  
are chosen as follows:  PSPS dihedrals are shown within  $[-240 ^{\circ} , 120 ^{\circ}]$,  SPSP dihedrals  are shown  within $[0 ^{\circ} , 360 ^{\circ}]$.}
\end{figure}

Besides, as more careful examination of the dihedral histograms  for both  (PSPS) and (SPSP) reveals, there exist some 
dihedral values (mainly in the kissing loops), that deviate strongly from the centres of the distributions 
(inset in  \ref{fig_3b_bad}). Their fraction is not high, however their presence has to be taken into account in a CG model.
The detailed dependences of the (PSPS) and (SPSP) dihedrals along the ring versus the dihedral index 
for ini2 SOP parameter set are shown in \ref{fig_dihfield} 
(Fig.~S5 show similar plots for ini1). In both cases, the  dihedrals deviating 
by $ \sim 180 ^\circ$ (we term them {\em cis})  from the distribution centres  are clearly visible
(the latter correspond approximately to  the  {\em trans} orientation). 
They belong to the 12 localized parts of the nucleic acid backbones 
(the sharp backbone kinks in \ref{fig_snaphair}) participating in 6 kissing loops, 
i.e. in total there are four such outstanding {\em cis} dihedrals per kissing loop pair. In the FF variant of the model, 
where all the dihedrals of the same type should have the same uniform equilibrium value, such dihedrals would be 
strongly discriminated against energetically by a harmonic or a quartic effective potential of Eq.~\ref{Veffquartic}, 
and this would strongly distort the equilibrium structure of the kissing loops. As a remedy, we
tested the following dihedral function:
\begin{equation}
V_{eff}(\phi) = \frac{k}{4} [1- \cos(2(\phi-\phi_0))], 
\label{Veffdihcos2}
\end{equation}
that has two minima  at $\phi = \phi_0$  and $\phi = \phi_0 + 180 ^\circ$, both with  stiffness $k$. 
It turns out that by accommodating the {\em cis} dihedrals, this simple function provides an excellent performance 
for the FF variant of the model in describing the RNA nanoring.

\begin{figure}
\includegraphics[width=3.25in]{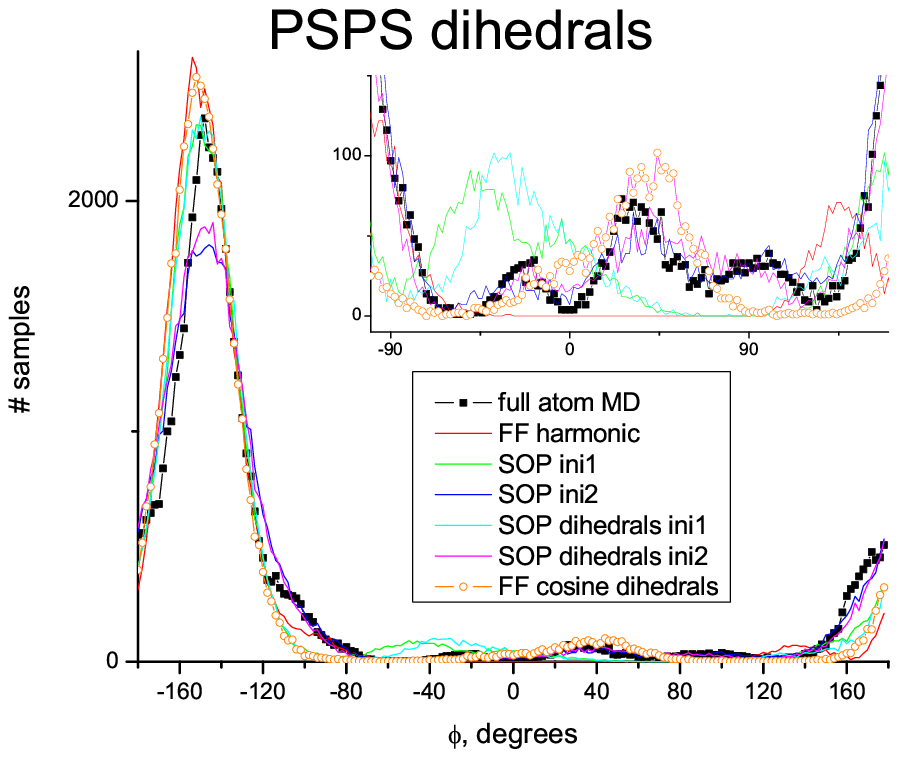}
\includegraphics[width=3.25in]{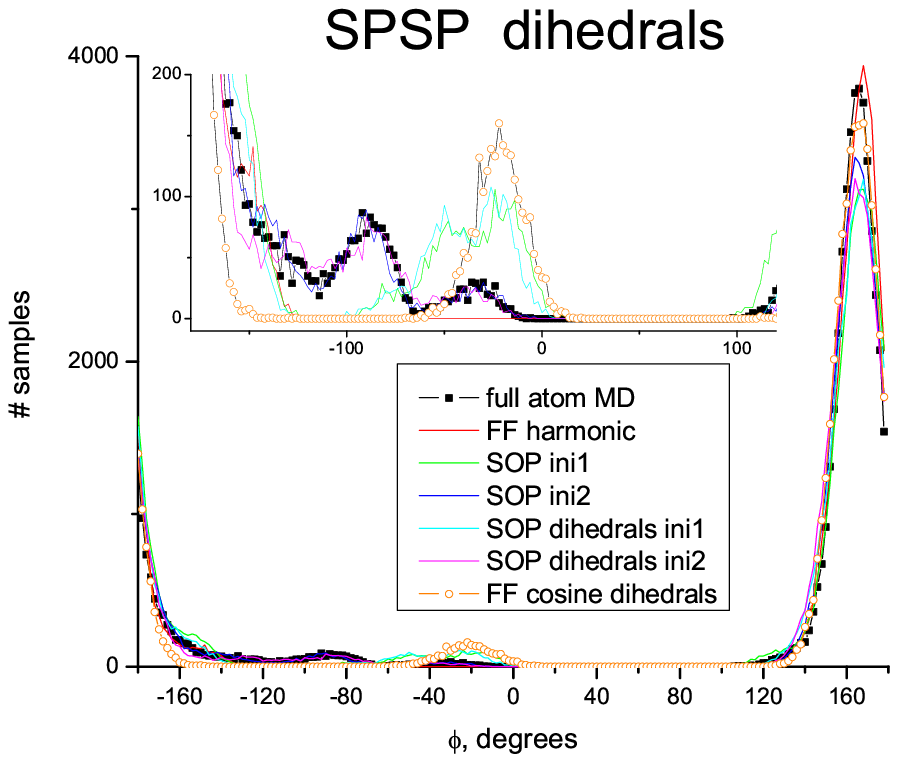}
\caption{\label{fig_cg_dihbp} 
Dihedrals  histograms for RNA nanoring from full atom MD and from CGMD runs for comparison.
The insets show the zoomed portions near the baselines  in the interval $[-100^\circ , 160^\circ]$ (PSPS)
and $[-180^\circ , 120^\circ]$ (SPSP), respectively. The small peaks near the baselines are 
reproduced equally well by SOP and SOP-dihedrals variants of the CG model, 
and they are also reasonably reproduced by FF-cosine-dihedrals variant, as  further discussed in the text.}
\end{figure}

\begin{figure}
\includegraphics[width=3.25in]{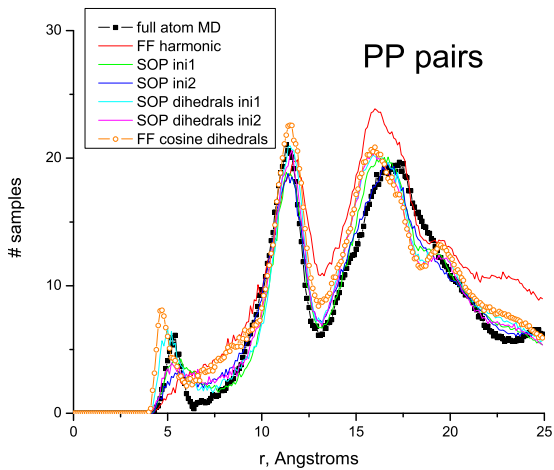}
\caption{\label{fig_cg_rdf} 
RDFs for nonbonded $PP$ pairs for the RNA nanoring from all-atom MD and from CGMD runs for comparison.}
\end{figure}

Since we intended to describe the RNA nanoring with a simple CG model based on the main body of purely helical parameters, 
with a controlled amount of non-helicity introduced either via SOP or via a FF with cosine dihedral term, 
we have chosen the set of  bonded parameters fitted via the BI method from 
the RNA dodecamer data, and tested it on the RNA nanoring. 
The \ref{tb:par3b} lists the values of parameters for the 3B CG model we thus obtained.
To  compare the performance of all the considered six variants of the 3B CG model, namely  the SOP, SOP-dihedrals 
(both with ini1 and ini2 detailed sets) as well as two FF variants 
with the harmonic Eq.~\ref{Veffquartic} (termed ''FF--harmonic'') and cosine Eq.~\ref{Veffdihcos2} 
(``FF--cosine--dihedrals'')  dihedral functions, we performed a series of $750$~ns long CGMD equilibration runs 
at constant temperature of $300$~K. The  dihedral histograms and RDFs  for nonbonded (PP) pairs obtained by the end of 
these runs  
are shown in \ref{fig_cg_dihbp} and \ref{fig_cg_rdf}, respectively, in comparison to the distributions from 
all-atom MD (full set of data can be found in the Figs.~S6-S10). After a few minor manual adjustments of the CG parameters, 
the histograms of the bonded terms show reasonable agreement with those from  the all-atom MD simulations 
(apart from some non-essential discrepancies further discussed in the remarks in Supporting Information). 
The dihedral distributions show the required extended tails in 
all cases except, obviously, FF--harmonic (\ref{fig_cg_dihbp}, the insets).   
What is even more remarkable, all the variants of the CG model (except the FF--harmonic) are able to capture the 
fine features of the nonbonded RDFs. The most important of them is the $5$~\AA \ peak for the $PP$ pairs 
(\ref{fig_cg_rdf}), which is even slightly over-emphasized by the FF--cosine--dihedrals variant. 

\begin{figure}
\includegraphics[width=2.5in]{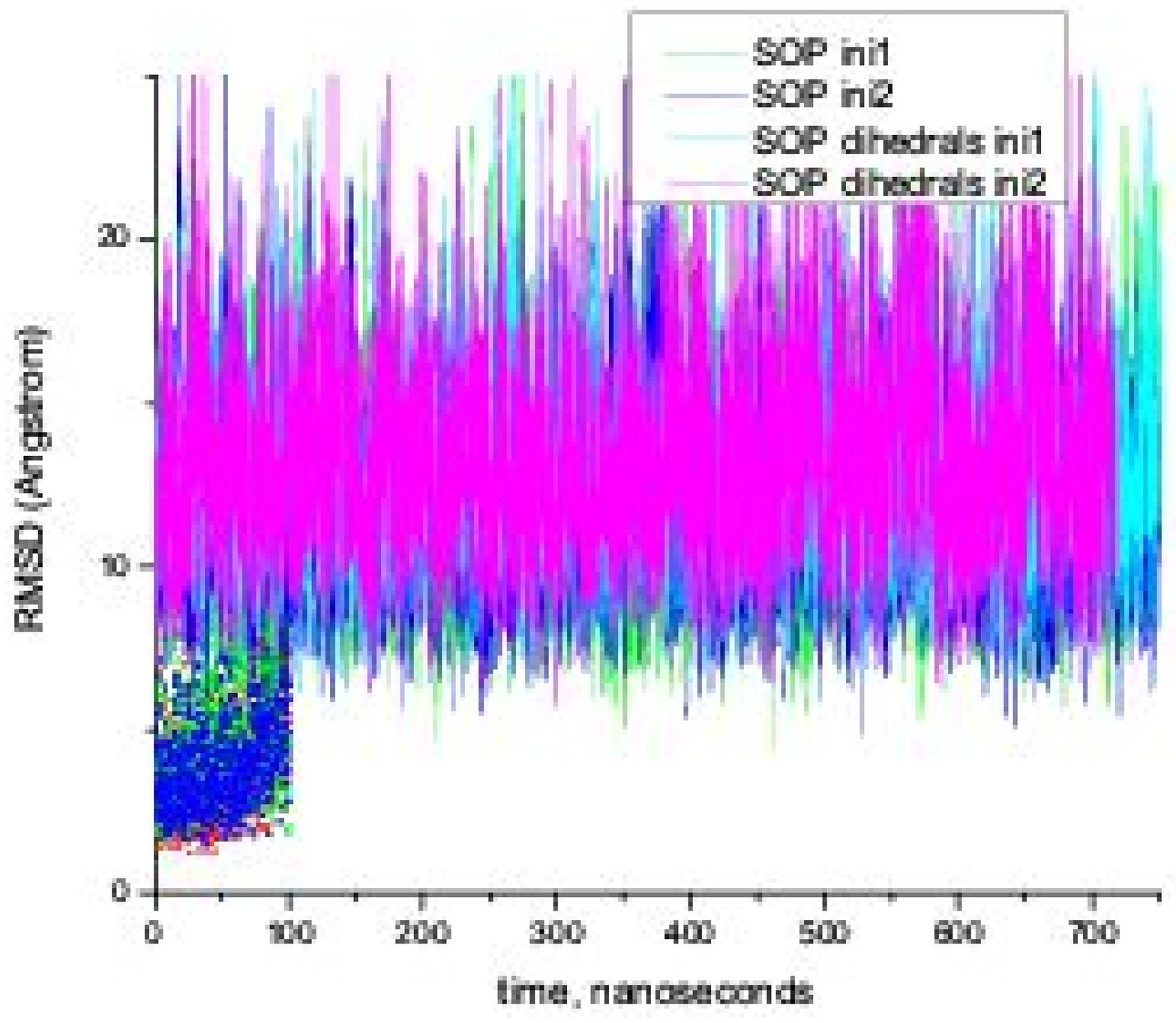}
\includegraphics[width=2.5in]{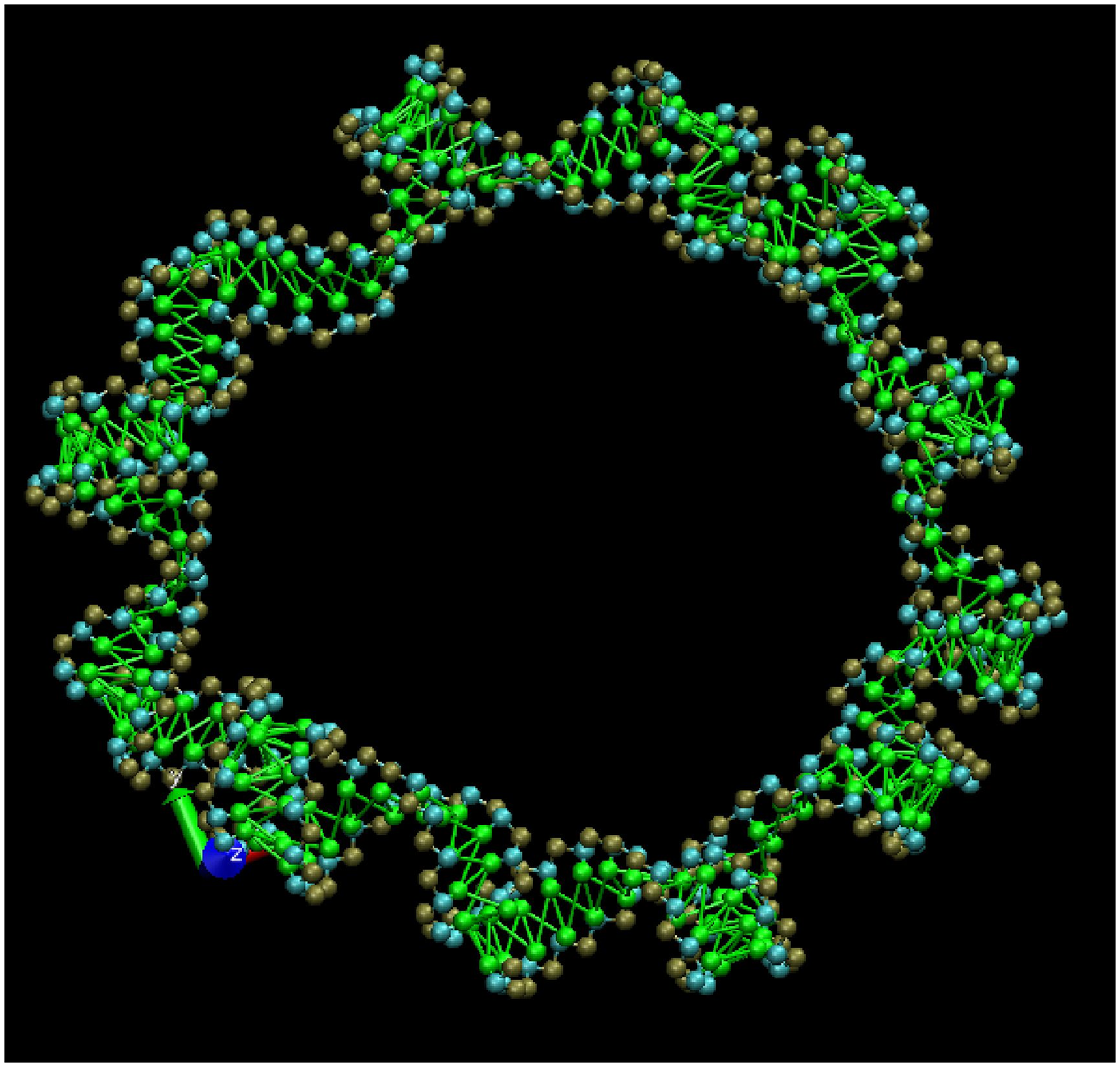}
\includegraphics[width=2.5in]{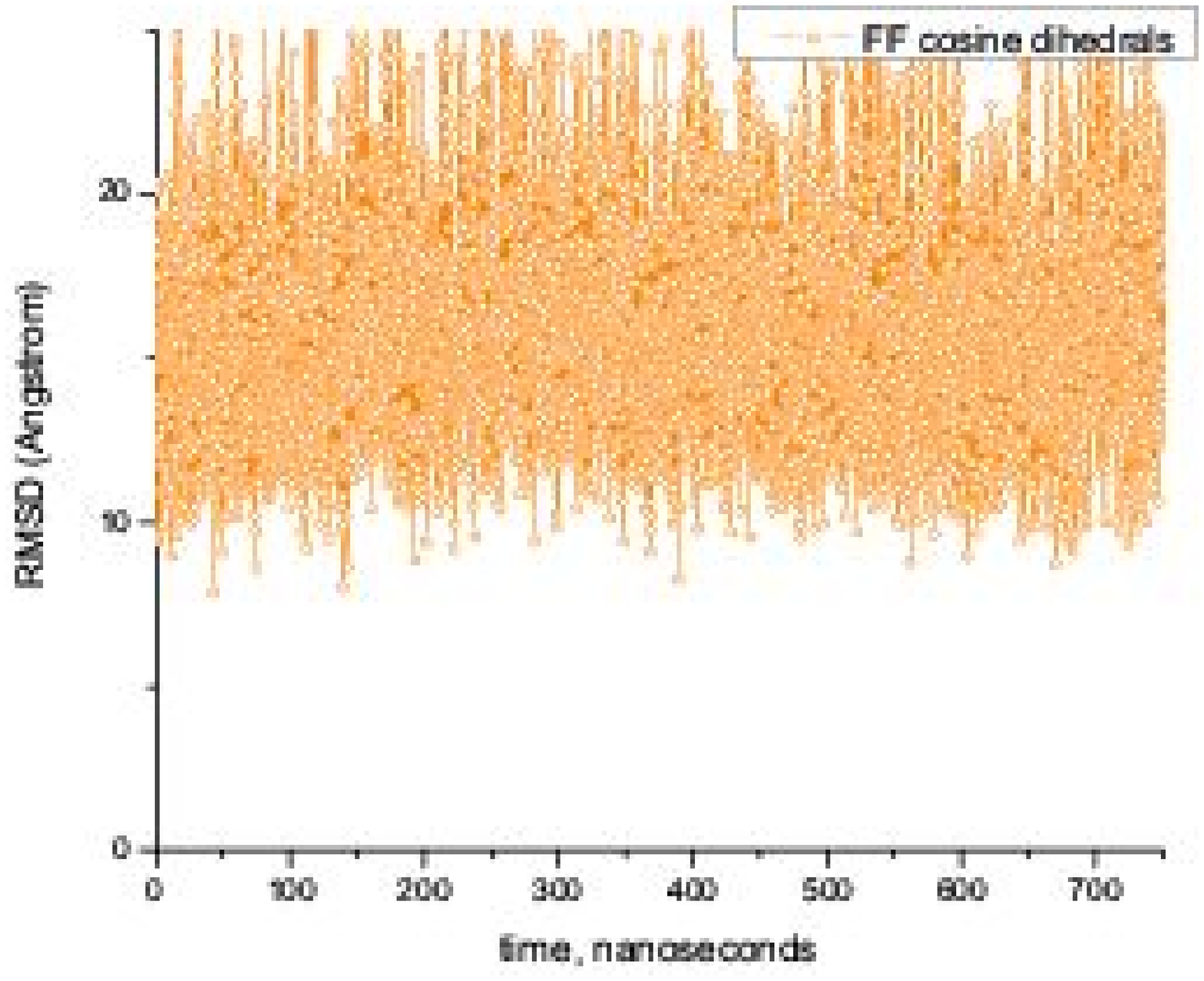}
\includegraphics[width=2.5in]{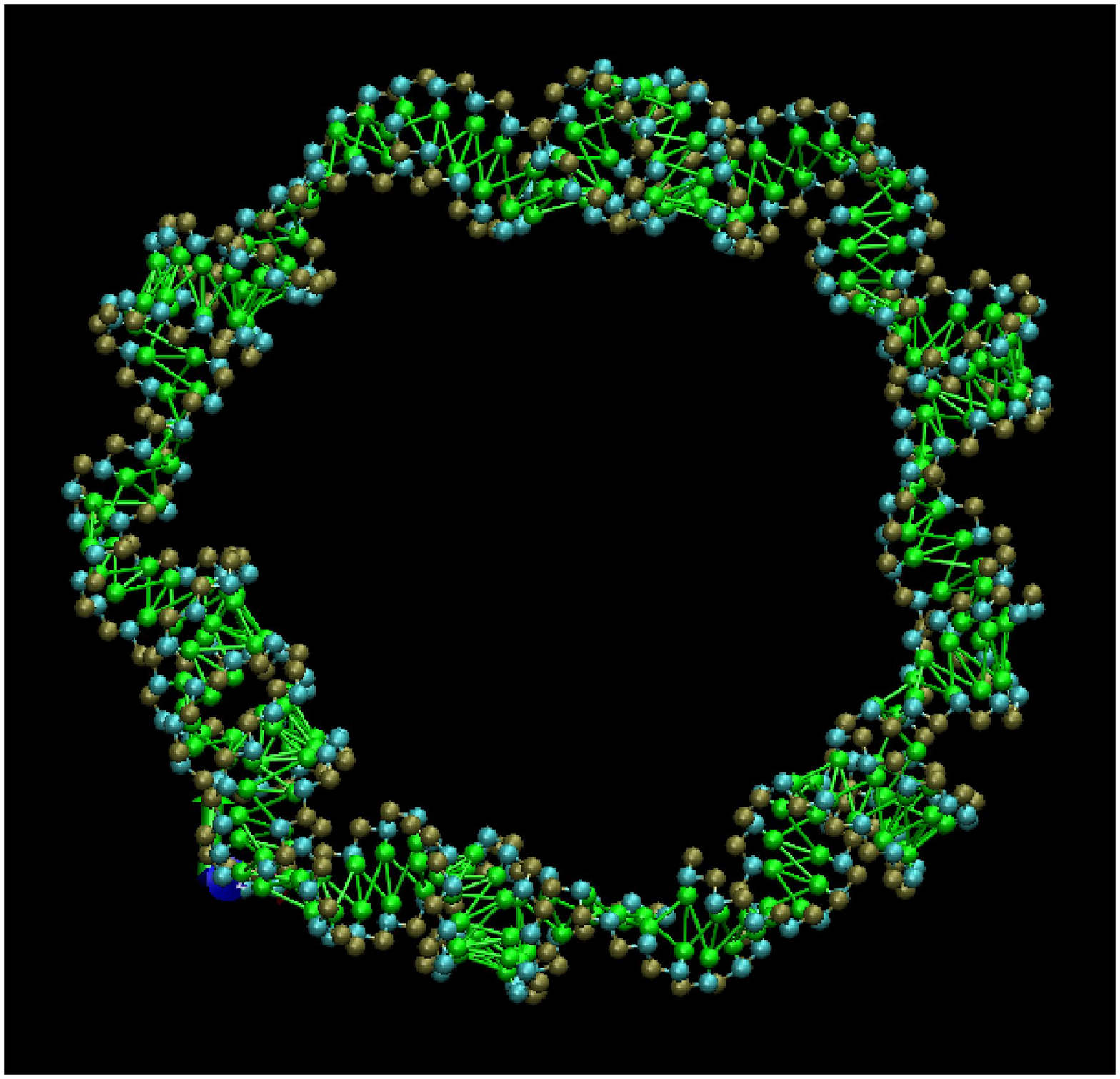}
\caption{\label{fig_rmsd} 
Left: RMSD for different variants of the parameter sets of the CG model in the equilibration CG runs. 
The data for the nanoring are plotted with lines and open circles ("FF cosine dihedrals"), 
while those for the dodecamer  are plotted with small symbols (in the interval up to $100$~ns only).
Right: The final snapshots of the RNA nanoring after  $750$~ns equilibration  in the CG model with the "SOP dihedrals" (top) and 
"FF cosine dihedrals" (bottom) parameter set.}
\end{figure}

The Root Mean Square Deviations (RMSDs) from the initial structures during these runs are plotted in \ref{fig_rmsd}.
Typical values of RMSD are  $ 13.0\pm 3.0$~\AA \ for SOP ini1 and ini2, $  13.5\pm 3.5$~\AA \ 
for SOP--dihedrals ini1 and ini2, and the RMSD is about the same ($  16.1\pm 3.2$~\AA ) for the  
FF--cosine--dihedrals variant. These values are to be 
compared  to the typical values for the dodecamer (also plotted in \ref{fig_rmsd} $  4.5\pm 1.5$~\AA \ 
for all considered variants) and to $ 35\pm 5.0$~\AA \ for the FF--harmonic variant  for  
the nanoring (not plotted) in which case the overall shape of the nanoring is not preserved. 
The final snapshots of the nanoring for the  SOP--dihedrals and FF--cosine--dihedrals variants are depicted in the same 
figure, attesting to the preservation of the helical segments, kissing loop structures and the overall shape. 

We conclude therefore, that  the 3B CG model provides an excellent description  for the structurally inhomogeneous 
RNA aggregate - the nanoring. This shows the power of the 3B representation, which, unlike the 1B one, 
captures adequately the excluded volume effects with small ($\approx 5$~\AA \ in size)  beads while allowing 
for closely spaced beads in the kissing loops.

\section{\label{discuss}Discussion: Coarse Grained Model and RNA conformation classes}

Two findings from the previous section are the most important. First, we observe that the SOP--dihedrals variants
 of the CG model provide about the same performance as the full SOP variants.
This means that structural complexity of the  RNA nanoring  can be handled by including into  a  3B CG model
the detailed information about P-C4' dihedral pseudo-angles only. 
This finding supports, for the case of the RNA nanoring, a more general statement, 
that  such a reduced representation of the RNA backbone gives a robust and complete description of the 
RNA structure \cite{RNAtorsional}, similar to the  famous $\phi - \psi$ Ramachandran plots for proteins 
[in the case of RNA the sugar pucker should be specified too \cite{RNAtorsional}, 
it is always C3'-{\em endo} in our case]. Moreover, based on a large body of  experimentally 
available RNA structures, it is shown in \cite{RNAtorsional}, that the values of  P-C4' dihedral pairs cluster in 
a few localised regions only in the 2D pseudo-torsional space, i.e. there exist quasi-discrete {\em RNA conformation classes}.

Second, in the light of the RNA conformation classes, another important finding is that a simple 
modification of the dihedral function Eq.~\ref{Veffdihcos2} allowed us to reach the 
SOP precision in describing the RNA nanoring by properly accommodating the distortions of the dihedrals in the kissing 
loops. Presently, the function Eq.~\ref{Veffdihcos2} has only one additional dihedral 
minimum separated from the original one by $180^\circ$,  i.e. only one 
additional conformation class (together with the main one for the A-form double helix). 
In principle, the dihedral function can be made more complex (e.g. to contain multiple minima). Note that  the 
number of observed conformation classes is limited by at most ten \cite{RNAtorsional}. 
Therefore, one can hope to keep the dihedral functions (for both PSPS and SPSP dihedrals) 
reasonably simple yet fairly universal, which opens up a promising  avenue towards a forcefield-like universal RNA CG model.

\begin{figure}
\includegraphics[width=3.25in]{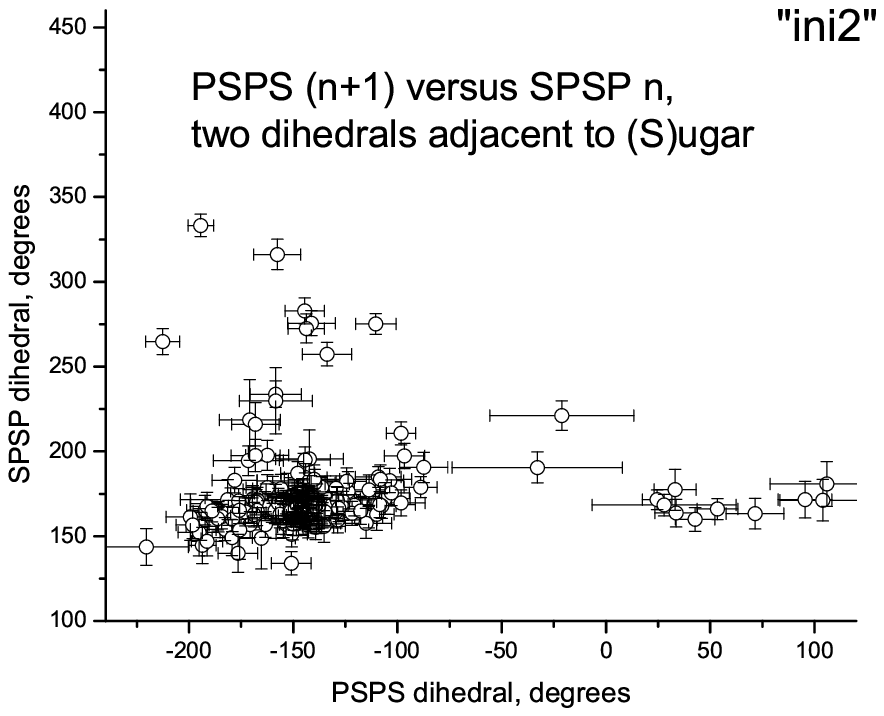}
\includegraphics[width=3.25in]{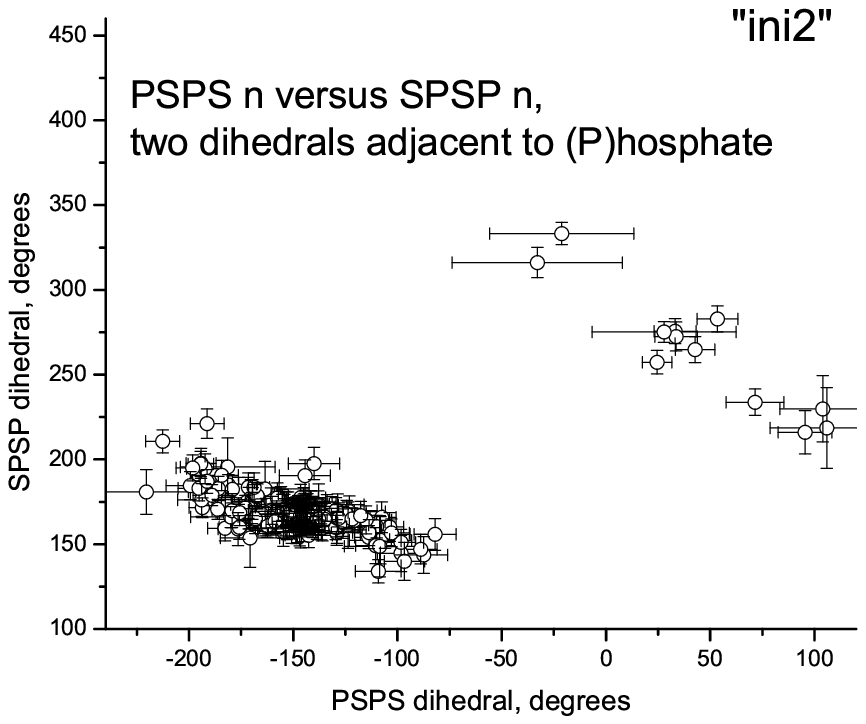}
\caption{\label{fig_dihmap} 
The 2D maps of the dihedral pair distributions of the RNA nanoring. 
Top: for the dihedral pairs centered around the sugars.
Bottom: for the dihedral pairs centered around the phosphates.}
\end{figure}

In this context it is interesting to establish a more detailed connection between the RNA conformation 
classes from \cite{RNAtorsional} and the ones that we observe in our simulations. 
According to \cite{RNAtorsional}, a conformation class is defined by a pair of  $(P_{i}S_{i}P_{i+1}S_{i+1})$ and $(S_{i-1}P_{i}S_{i}P_{i+1})$
dihedrals centered around a sugar $S_i$. Note that alternatively one can also select dihedral pairs centered around a 
phosphate $P_i$, i.e.  $(S_{i-1}P_{i}S_{i}P_{i+1})$ and $(P_{i-1}S_{i-1}P_{i}S_{i})$.
Using the RNA nanoring dihedrals from \ref{fig_dihfield}, we plotted  2D maps for both variants (\ref{fig_dihmap}). 
The main double-helical peak at $(-153^\circ , 169^\circ)$ is clearly visible, 
along with the two shoulders  extending in the PSPS direction (cf. \ref{fig_3b_bad}).
In the terminology of \cite{RNAtorsional} the shoulders correspond to the classes VI (cross-stem stacking of the purine-purine base pairs) and IV  
(absent stacking on the 3' side of the nucleotide). The former is present in the nanoring 
because of a cross-stem stacking G-G pair just near the base of each loop \cite{LeeCrothers}, the 
latter is probably found in the middle of the nanoring sides (we did not analyse this in detail). 

Two more classes are found in the kissing loops (\ref{fig_dihmap}, top). 
Namely, there are (i) 12 nucleotides (one per each kissing loop side) 
that have strongly deviating {\em cis} PSPS angles and double-helical {\em trans} SPSP angles, and 
(ii) immediately following them in the 5' to 3' sense, 12 other nucleotides that have the situation reversed, 
i.e. {\em cis} SPSP angles and  {\em trans} PSPS angles. These nucleotides (found near  sharp kinks of the nucleic 
backbone visible in \ref{fig_snaphair})  are the last ones base paired within their 
own chain and the first ones participating in the cross--chain base pairing in the kissing loops, 
respectively. While we can relate our class (ii) to the class II from \cite{RNAtorsional}, our class (i) seems to 
be absent from the scheme of \cite{RNAtorsional}.   
Interestingly, two classes (i) and (ii)  collapse into a single one if the 2D dihedral map is replotted with
the dihedral pairs centered around the phosphates (\ref{fig_dihmap}, bottom). The reason for this is clear for the case 
of the nanoring, where the nucleotides of the classes (i) and (ii) are neighbours in the chain, 
but one may wonder whether a similar procedure applied to all the volume of data analysed in \cite{RNAtorsional} 
would lead to a simplification of the  observed conformation classes. Note, that the classification based 
on the phosphates (``suites''), and not on the nucleotides has  already been used for the full RNA
conformational space \cite{Murray2003}.

Thus, we incorporated only the classes (i) and (ii) [collapsing to a single class if reformulated as specified above] 
into the FF--cosine--dihedrals variant of our CG model, and ignored the remaining two classes present in the nanoring 
[IV and VI according to  \cite{RNAtorsional}]. All these classes are obviously accounted for in the SOP--dihedrals variant. 
This explains slightly elevated values of RMSD in the former case, compared to the latter, though 
one extra class alone proved to be sufficient to model the kissing loops reasonably well.

\section{\label{concl}Conclusions}

In the present paper we reported  a series of bead-based CG models for RNA with varying  amounts of atomistic structural 
information and  numbers of beads per nucleotide. They range from SOP ones, where all the specific values 
of bonds/angles/dihedrals from a reference structure (nanoring) are included in the model,  
to a forcefield approach, where the CG model is described by a few universal parameters. We started from the purely double-helical 
set of parameters derived from  all-atom MD data for an RNA dodecamer, and applied it to the RNA nanoring, introducing the non-helical 
features whenever necessary. We are concluding here that the models with just one bead per nucleotide suffer from the 
effects of insufficient excluded volume, while the models with three beads  per nucleotide [(P)hosphate, (S)ugar, (B)ase] are more
 suitable for the description of the RNA. The inclusion of just the detailed information about the (PSPS) and (SPSP) dihedral  angles in the model 
renders precision  similar to the inclusion of all available atomistic structural information, which illustrates  the
usefulness and robustness of the reduced (P-C4') representation of the nucleic backbone \cite{RNAtorsional}. 
Furthermore, the existence of the quasi-discrete RNA conformation classes based on these dihedrals  \cite{RNAtorsional}
is supported by our data too. For the simple non-helical conformation of the kissing loops we were able to 
design a dihedral potential function Eq.~\ref{Veffdihcos2}, that, while including only a single additional minimum, 
successfully accommodated local dihedral distortions. This opens up the road towards the development of
even more transferable RNA CG models based on the P-C4' conformation classes.  \ref{tb:par3b}  lists the values of  the parameters ($26$ in total) for the 3B CG model we obtained.
For the systems of thousands of nucleotides, a time scale of microseconds can be easily reached with the developed CG model. 
The structural precision of our 3B models in terms of RMSD is $\sim 0.06 $~\AA \ per nucleotide 
[to be compared  e.g. with  $\sim 0.1 $~\AA \ for another recent 3B model \cite{DokholyanRNA} and  $\sim 0.13 $~\AA \ 
for a 1B model \cite{Jonikas}].

Finally, we mention several directions for future development of our RNA CG model. 
Explicit electrostatics between phosphates as well as the sequence-specificity in 
the base-pair interactions will be incorporated. Besides, if the interaction between base pairs were treated in a non-bonding 
manner, this would allow one to study the association/dissociation reactions between the  building blocks of the nanoring 
and other RNA nanostructures.

\section{\label{suppinfo}Supplemental information}

The following extras can be of interest for some readers and is available 
on request: (i) details of the all-atom MD simulations used as sources for CG model fitting;
(ii) details of the 1B CG  model simulations; (iii) full set of graphs for all CG degrees of freedom.

\begin{acknowledgements}
M.P.\ and R.M.\ are grateful to the NSERC CRC Program for its support. 
B.A.S.\ was supported in part by the intramural research program of the NIH, 
National Cancer Institute, Center for Cancer Research. 
This work was made possible by the facilities of the Shared Hierarchical
Academic Research Computing Network (SHARCNET:www.sharcnet.ca). 
The authors are grateful to Valentina Tozzini for helpful comments and suggestions.
\end{acknowledgements}

\begin{table}
\begin{center}
\begin{tabular}{l l l}
  \hline 
  \hline 
  \multicolumn{3}{c}{3B CG model parameters} \\
  \hline 
  \hline 
  \multicolumn{3}{c}{Backbone bonds} \\
  \hline  
    & $r_0$  & $k$   \\
  \hline  
  $(P_{i}S_{i})$ & 3.93 & 133.4  \\
  \hline  
  $(S_{i}P_{i+1})$ & 3.92 & 107.8   \\
  \hline  
  $(S_{i}B_{i})$ & 3.38 & 61.3   \\
  \hline
  \hline 
  \multicolumn{3}{c}{Base pairing bonds} \\
  \hline  
    & $r_0$ & $k$    \\
  \hline  
  $(B_{i}B_{j})$  & 8.99 & 16.5   \\
  \hline  
  $(B_{i+1}B_{j})$  & 6.86 & 2.83   \\
  \hline  
  $(B_{i+2}B_{j})$ & 7.35 & 2.93   \\
  \hline  
  \hline 
  \multicolumn{3}{c}{Backbone angles} \\
  \hline  
    & $\theta_0$ & $k$  \\
  \hline  
  $(P_{i}S_{i}P_{i+1})$ & 1.733 ($99.3^{\circ}$) & 26.3  \\
  \hline  
  $(S_{i}P_{i+1}S_{i+1})$ & 1.819 ($104.2^{\circ}$) & 106.9   \\
  \hline  
  $(P_{i}S_{i}B_{i})$ & 1.728 ($99.0^{\circ}$) & 32.1   \\
  \hline  
  $(B_{i}S_{i}P_{i+1})$ & 1.642 ($94.1^{\circ}$) &  127.2  \\
  \hline
  \hline 
  \multicolumn{3}{c}{Backbone dihedrals} \\
  \hline  
    & $\phi_0$  & $k$    \\
  \hline  
  $(P_{i}S_{i}P_{i+1}S_{i+1})$ & -2.677 ($-153.4^{\circ}$) & 5.86   \\
  \hline  
  $(S_{i}P_{i+1}S_{i+1}P_{i+2})$ & 2.948 ($168.9^{\circ}$) & 16.2  \\
  \hline  
  \hline  
  \multicolumn{3}{c}{WCA potential for nonbonded pairs} \\
  \hline  
    & $\sigma$  & $\varepsilon$     \\
  \hline  
  all & 5.0  & 0.1    \\
  \hline  
  \hline  
  \multicolumn{3}{c}{Excluded bonds (see the text)} \\
  \hline  
$(S_{i}B_{i+1})$ & $(B_{i}S_{i+1})$ & $(B_{i}B_{i+1})$  \\
  \hline  
  \hline  
\end{tabular}
\end{center}
\caption{Parameters of 3B  CG model.  The distances are expressed in \AA , the angles and dihedrals are expressed in radians 
(the values in degrees are shown for convenience too). The unit of energy is kcal/mol, and the units for the 
coefficients are derived from the unit of energy and  \AA \ or radians respectively. 
The  nucleotide index $i$ is counted from the 5' end.}
\label{tb:par3b}
\end{table}

\bibliographystyle{apsrev}

\bibliography{C:/1Articles/rna}

\end{document}